\documentclass[prl,aps,twocolumn,superscriptaddress,showpacs,nofootinbib]{revtex4-2}

\usepackage{graphicx}
\usepackage{textcomp}
\usepackage{dcolumn}
\usepackage{amsmath}
\usepackage{siunitx}
\usepackage{booktabs}
\usepackage{array}
\usepackage[colorlinks]{hyperref}
\hypersetup{
	colorlinks = true,
	citecolor = {blue},
	linkcolor = {blue},
	urlcolor = {blue}
}
\begin{document}

\title{High-Performance Quantum Frequency Conversion from Ultraviolet to Telecom Band}

\author{Yi Yang}
\affiliation{Hefei National Research Center for Physical Sciences at the Microscale and School of Physical Sciences, University of Science and Technology of China, Hefei 230026, China.}
\affiliation{Jinan Institute of Quantum Technology and CAS Center for Excellence in Quantum Information and Quantum Physics, University of Science and Technology of China, Jinan 250101, China.}

\author{Bin Wang}
\affiliation{Jinan Institute of Quantum Technology and CAS Center for Excellence in Quantum Information and Quantum Physics, University of Science and Technology of China, Jinan 250101, China.}

\author{Ji-Chao Lin}
\author{Yang Gao}
\author{Xin Li}
\affiliation{Jinan Institute of Quantum Technology and CAS Center for Excellence in Quantum Information and Quantum Physics, University of Science and Technology of China, Jinan 250101, China.}

\author{Jiu-Peng Chen}
\affiliation{Hefei National Laboratory, University of Science and Technology of China, Hefei 230088, China.}
\affiliation{Jinan Institute of Quantum Technology and CAS Center for Excellence in Quantum Information and Quantum Physics, University of Science and Technology of China, Jinan 250101, China.}

\author{Lei Hou}
\affiliation{Hefei National Laboratory, University of Science and Technology of China, Hefei 230088, China.}
\affiliation{Shanghai Research Center for Quantum Science and CAS Center for Excellence in Quantum Information and Quantum Physics, University of Science and Technology of China, Shanghai 201315, China.}

\author{Ye Wang}
\author{Yong Wan}
\affiliation{Hefei National Research Center for Physical Sciences at the Microscale and School of Physical Sciences, University of Science and Technology of China, Hefei 230026, China.}
\affiliation{Hefei National Laboratory, University of Science and Technology of China, Hefei 230088, China.}
\affiliation{Shanghai Research Center for Quantum Science and CAS Center for Excellence in Quantum Information and Quantum Physics, University of Science and Technology of China, Shanghai 201315, China.}

\author{Xiu-Ping Xie}
\author{Ming-Yang Zheng}
\affiliation{Jinan Institute of Quantum Technology and CAS Center for Excellence in Quantum Information and Quantum Physics, University of Science and Technology of China, Jinan 250101, China.}
\affiliation{Hefei National Laboratory, University of Science and Technology of China, Hefei 230088, China.}

\author{Qiang Zhang}
\affiliation{Hefei National Research Center for Physical Sciences at the Microscale and School of Physical Sciences, University of Science and Technology of China, Hefei 230026, China.}
\affiliation{Jinan Institute of Quantum Technology and CAS Center for Excellence in Quantum Information and Quantum Physics, University of Science and Technology of China, Jinan 250101, China.}
\affiliation{Hefei National Laboratory, University of Science and Technology of China, Hefei 230088, China.}
\affiliation{Shanghai Research Center for Quantum Science and CAS Center for Excellence in Quantum Information and Quantum Physics, University of Science and Technology of China, Shanghai 201315, China.}

\author{Jian-Wei Pan}
\affiliation{Hefei National Research Center for Physical Sciences at the Microscale and School of Physical Sciences, University of Science and Technology of China, Hefei 230026, China.}
\affiliation{Hefei National Laboratory, University of Science and Technology of China, Hefei 230088, China.}
\affiliation{Shanghai Research Center for Quantum Science and CAS Center for Excellence in Quantum Information and Quantum Physics, University of Science and Technology of China, Shanghai 201315, China.}

\date{\today}

\begin{abstract}
Quantum frequency conversion (QFC) is essential for bridging the spectral gap between stationary qubits and low-loss optical communication channels. 
In this work, we demonstrate a short-wavelength-pumping QFC with the first-order quasi-phase matching period of $\SI{3.07}{\micro\meter}$ on thin-film lithium niobate, converting ultraviolet photons to the telecom C-band. By constructing a theoretical model that correlates the normalized conversion efficiency with domain defects in the short-period phase-matched waveguide, we found the critical tolerance of domain defects along the waveguide should be $\le 2$ (excluding the ends). Based on this, we achieved a theoretical limit normalized conversion efficiency of 839\%/(W·c\(\mathrm{m^{2}}\)) for the fundamental guided mode through fabrication optimization. Furthermore, we propose a robust noise suppression strategy for short-wavelength pumping by utilizing the counter-tuning behaviors of difference-frequency generation and spontaneous parametric down-conversion. By combining these advances with ultra-narrowband filtering, we achieve a record-high external efficiency of 28.8\% and an ultra-low noise of 35 counts per second. This high-performance QFC connecting ultraviolet and telecom bands satisfies the stringent requirements for long-lived remote ion-ion entanglement in scalable quantum networks [W.-Z. Liu \textit{et al.}, Nature (2026)].




\end{abstract}

\maketitle
\textbf{\textit{Introduction.}}---
Quantum networks promise revolutionary applications such as distributed quantum computing, secure communication, and quantum-enhanced sensing~\cite{kimble2008Nature,popkin2021-science-quantum-internet}. Realizing such networks requires distributing entanglement over long distances, fundamentally limited by fiber loss~\cite{briegel1998PRL-repeaters}. The majority of early demonstrations with atomic ensembles~\cite{kimble2005-nature-atomic-ensembles,kuzmich2005-nature-atomic-ensembles,yuan2008nature-atomic-ensembles}, single neutral atoms~\cite{ritter2012Nature-atoms}, trapped ions~\cite{moehring2007Nature-iontrap-369-1m,Oxford2022Nature-Sr488-2m}, and nitrogen-vacancy centers~\cite{pompili2021Science-NV-centers} have been constrained to meter-scale distances due to severe fiber attenuation at their native wavelengths. Quantum frequency conversion (QFC) overcomes this barrier by translating photons to low-loss telecom bands while preserving quantum coherence~\cite{kumar1990-ol-qfc-qfc-theory,huang1992-prl-qfc-exp,ikuta2011nc-preserve-coherence,zaske2012rpl-visible-qfc-preserve-coherence}, thereby facilitating long-distance and hybrid quantum interconnects.


Recently, efficient and low-noise QFC interfaces based on difference-frequency generation (DFG) have been realized for various quantum memories operating in the visible and near-infrared regimes, including neutral atoms~\cite{bao2020nature-795->1342-22km,bao2022prl-795->1342-20.5km,Weinfurter2022Nature-atom-780-33km,weinfurter2024-prxquantum-780->1517-101km,bao2024nature-795->1342-12.5km,2504.05660}, trapped ions~\cite{walker2018PRL-iontrap-866-10km,krutyanskiy2019NPJ-854-50km,krutyanskiy2024prx-iontrap-854-101km}, and solid-state color centers~\cite{hanson2024-637->1588-nv-25km,bersin2024telecom-737nm-QFC,knaut2024entanglement-SiV-737nm}, enabling entanglement distribution over fiber links spanning tens to hundreds of kilometers. Trapped-ions systems with dominant emission in the ultraviolet (UV) and blue spectral region possess exceptional properties including long coherence times~\cite{wang2021-long-coherence-time-1.5h,mjqd-9mvf-long-coherence-time-2h} and high-fidelity control~\cite{harty2014-high-fidelity-single-qubit-99.9999,ballance2016-high-fidelity-two-qubit-99.9,srinivas2021-high-fidelity-two-qubit-99.92,smith2025-high-fidelity-single-qubit-99.99999}, positioning them as prime candidates for scalable quantum networks. Nevertheless, networks relying on trapped ions are typically restricted to meter-scale distances~\cite{moehring2007Nature-iontrap-369-1m, Oxford2022Nature-Sr488-2m}, primarily due to the poor QFC performance in the short-wavelength band.


Among the prior efforts~\cite{kasture2016JOP-UVQFC-Yb369->1580,wright2018PRA-UVQFC-Sr425->1560-1548->422,yu2025APL-UVQFC-Yb369->456->864->1552} focusing on QFC down-converting short wavelengths to telecom bands, nonlinear processes with higher-order quasi-phase matching (QPM) or multi-stage DFG are employed. Higher-order QPM relaxes the stringent requirements for high-quality short-period poling but suffers from low normalized conversion efficiency~\cite{kasture2016JOP-UVQFC-Yb369->1580}. Multi-stage DFG, using long-wavelength pumping to mitigate pump-induced noise, introduces substantial optical losses and therefore yields low system efficiency~\cite{yu2025APL-UVQFC-Yb369->456->864->1552}. Overall, these approaches have achieved only a few percent system efficiency with noise levels of several thousand counts per second (cps), far below practical requirements of remote entanglement in scalable quantum networks~\cite{liu2026-393nature}. 


\begin{figure*}[htbp]
	\centering
	\includegraphics[width=1\linewidth]{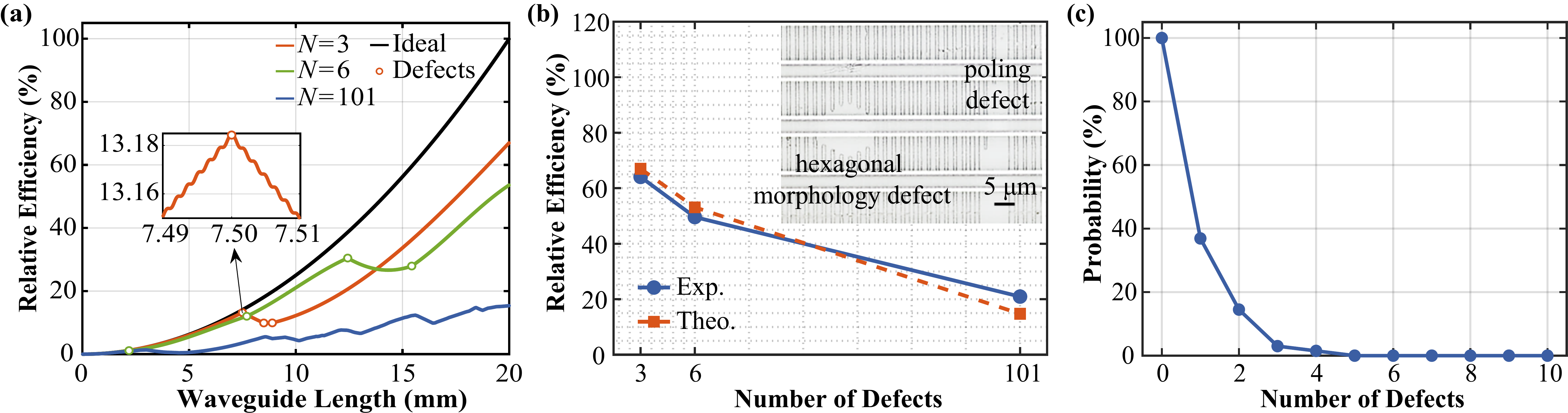}
        \caption{(a) Evolution of the relative efficiency along the waveguide for $N=3$ (orange), 6 (green), and 101 (blue) defects. The black curve represents the ideal growth without defects. Open circles mark the positions where domain defects induce abrupt transitions in efficiency evolution. Inset: highlights the local perturbation at a specific defect site. (b) Comparison of relative efficiency between theory (blue solid line) and experiment (orange dashed line) based on measured defect statistics. Inset: displays a micrograph of the domain structure. (c) Simulated probability of the normalized efficiency exceeding 90\% of its ideal value as a function of defect count.}
        \label{Fig:group_figure1}
\end{figure*}
In this work, we demonstrate a high-performance QFC interface, converting 393 nm photons from $^{40}$Ca$^+$ ions to the telecom C-band at 1550 nm. Through quantitatively modeling the impact of domain defects on normalized conversion efficiency ($\eta_{\rm nor}$) and optimizing the periodical poling process, an ultra-high $\eta_{\rm nor}$ of 839\%/(W·c\(\mathrm{m^{2}}\)) is achieved in a periodically poled lithium niobate (PPLN) ridge waveguide, which lowers the pump power required for maximum conversion efficiency and consequently suppresses the pump-induced nonlinear noises. Furthermore, utilizing the counter-tuning characteristics of the DFG and spontaneous parametric down-conversion (SPDC) noises for short-wavelength pumping, roughly a threefold reduction in noise count rate is achieved. Based on these optimizations, the QFC module achieves an external efficiency of 28.8\%, with a noise level of 35 cps, surpassing previous UV/blue-to-telecom reports~\cite{kasture2016JOP-UVQFC-Yb369->1580,wright2018PRA-UVQFC-Sr425->1560-1548->422,yu2025APL-UVQFC-Yb369->456->864->1552} by over 30-fold in efficiency and more than two orders of magnitude in noise reduction. Crucially, the device reported here supports a signal-to-noise ratio of $\sim$ 120:1 for trapped $^{40}\text{Ca}^{+}$ ions, enabling the recent demonstration of device-independent quantum key distribution over long-distance fiber links~\cite{liu2026-393nature}.



\textbf{\textit{Theoretical-Limit QPM Process.}}---The normalized efficiency reported in most frequency conversion studies based on ridge waveguides is typically limited to 30\%--70\% of the theoretical value~\cite{lkrutyanskiy2017apb-854to1550,walker2018PRL-iontrap-866-10km,bock2018-natcommun-ca854->1310,weinfurter2020prl-atom-780-20km}. This challenge is exacerbated in UV-to-telecom conversion by stringent short-period poling requirements~\cite{kasture2016JOP-UVQFC-Yb369->1580}. For PPLN waveguides fabricated on lithium niobate film with a thickness of micrometers, the upper limit of $\eta_{\rm nor}$ is primarily determined by the quality of domain inversion rather than thickness uniformity. Defects (Fig.~\ref{Fig:group_figure1}(b)) arising from intrinsic lattice pinning (hexagonal morphology defect) and waveguide fabrication (poling defect) induce phase mismatches, disrupting coherent accumulation and reducing conversion efficiency.

To determine the correlation between domain defects and the theoretical $\eta_{\rm nor}$, we quantitatively modeled the impact of random domain defects on normalized conversion efficiency. The normalized conversion efficiency is governed by the Fourier transform of the nonlinear coefficient distribution $d(z)$, given by $\eta \propto \left| \int_{0}^{L} d(z) e^{-i 2\pi q z} dz \right|^2$. Specifically, for a waveguide of length $L$ containing $N$ randomly distributed domain defects, $d(z)$ can be formulated as~\cite{boyd2008nonlinear,hariharasubramani2013-defect-analysis}:
\begin{equation}
d(z) = d_{\text{eff}} \sin\left(\frac{2\pi z}{\Lambda}\right) \cdot \exp\left[i \sum_{j=0}^{N} \left(\phi_j*H(z-x_j)\right)\right],
\label{Eq:dz}
\end{equation}
with
\begin{equation}
\phi_j = \frac{2\pi}{\Lambda} \left( w_j - \frac{\Lambda}{2} \right), \quad H(z-x_j) = \begin{cases} 1, & z \ge x_j
 \\ 0, & z < x_j \end{cases},
\label{Eq:dz}
\end{equation}
where, $\phi_j$ is the phase shift introduced by the $j$-th domain defect, $w_j$ is the defect width, and $x_j$ is the starting position of defect ($x_0 = 0$), respectively. Incorporating measured defect statistics into this model, we simulate the efficiency evolution along the waveguide (Fig.~\ref{Fig:group_figure1}(a)), illustrating how defect-induced phase shifts disrupt the ideal QPM process and degrade efficiency. The close agreement between predicted relative efficiencies and experimental values (Fig.~\ref{Fig:group_figure1}(b)) confirms the validity of our model. Building on this, we simulate the probability of $\eta_{\rm nor}$ exceeding 90\% of its ideal value as a function of defect count for a 20-mm-long waveguide (Fig.~\ref{Fig:group_figure1}(c)), under the assumption that defect positions are randomly distributed and their widths follow an experimentally derived Poisson distribution ($\lambda$ $\sim$ \SI{12.3}{\micro\meter}). Notably, with merely two defects, the probability drops to approximately 20\%, implying that a defect count of $\le 2$ is required for near-theoretical performance (see Supplemental Material for more details).



\begin{figure}[htbp]
	\centering
	\includegraphics[width=1\linewidth]{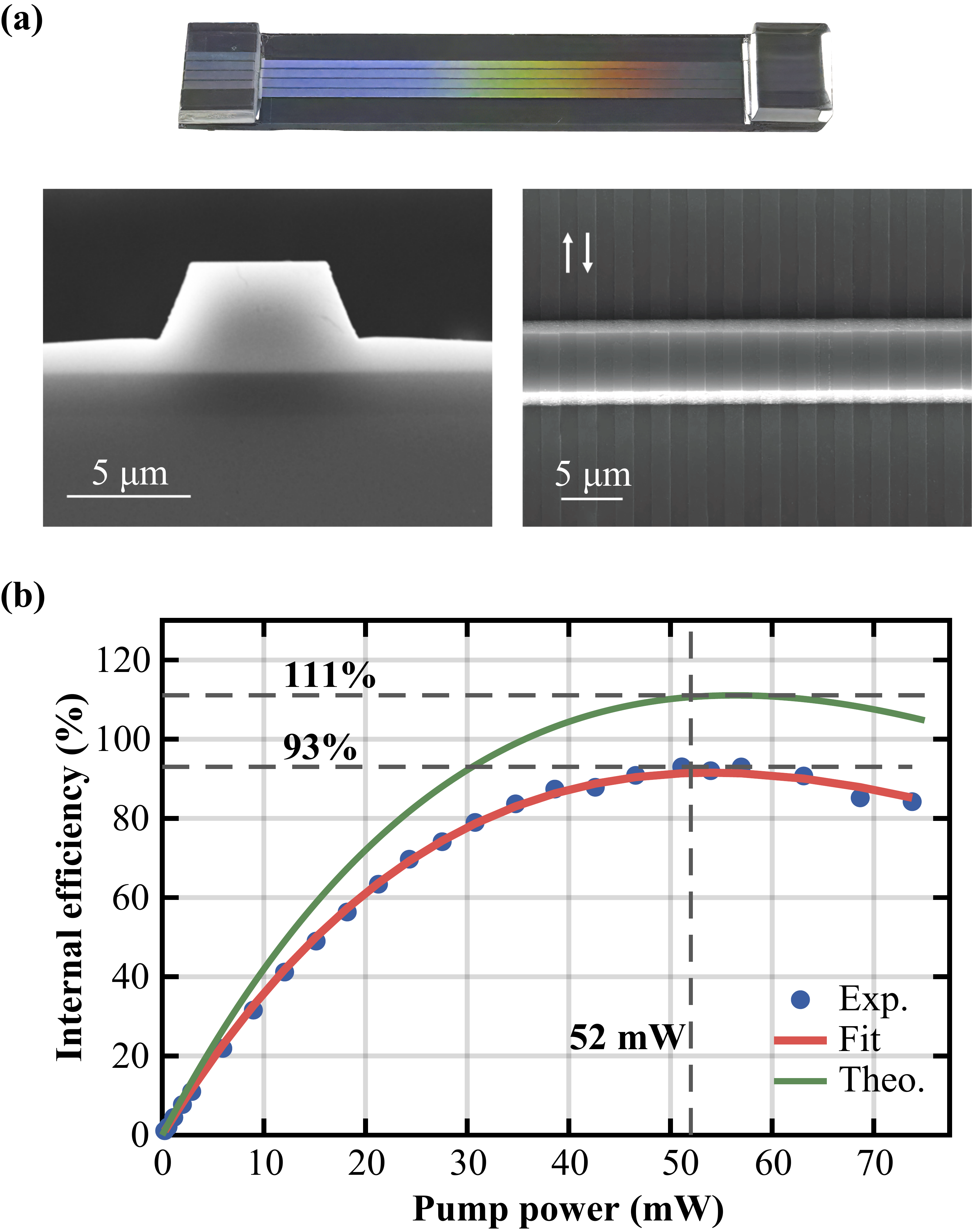}
        \caption{(a) Images of the fabricated device: (Top) Photograph of the fabricated ridge waveguide chip; (Bottom left) Scanning electron microscopy (SEM) image of the waveguide facet; (Bottom right) SEM image of the periodically poled domain structure revealed by hydrofluoric acid etching. (b) Internal efficiency versus pump power. Filled blue circles: measured points; solid red line: $\sin^2$-type fit to the experimental data; solid green line: the theoretical prediction from the coupled-wave equations including propagation losses.}
        \label{Fig:group_figure2}
\end{figure}

To meet the stringent domain defect threshold, optimizing the waveguide fabrication process is essential. We designed a 20-mm-long ridge waveguide for down-converting 393 nm photons to the 1550 nm telecom band (Fig.~\ref{Fig:group_figure2}(a)). The device is fabricated on a z-cut MgO-doped thin-film LN platform comprising a \SI{5}{\micro\meter} LN layer on \SI{2}{\micro\meter} buried silicon oxide. The ridge is defined by inductively coupled plasma etching with a top width of \SI{5}{\micro\meter} and ridge height of \SI{3.5}{\micro\meter}. QPM is achieved via periodic poling with $\Lambda = \SI{3.07}{\micro\meter}$ for first-order type-0 interaction. Critically, domain defect control in this workflow relies on two pivotal measures: the pre-selection of wafers with minimal intrinsic pinning-induced hexagonal morphology defect counts, and the mitigation of poling defects arising from photolithographic patterning distortions, thermal instabilities, or mechanical damage.


To evaluate the efficacy of the defect control measures and verify the theoretical limit, we derive the experimental $\eta_{\rm nor}$ by characterizing the internal conversion efficiency ($\eta_{\rm int}$). Here, $\eta_{\rm int}$ is defined in terms of the powers measured at the waveguide output ($\eta_{\rm int} \equiv \lambda_3 P_3^{\rm out}/\lambda_1 P_1^{\rm out}$) and is measured as a function of pump power ($P_2$), where subscripts 1, 2, and 3 denote signal, pump, and DFG, respectively. Experimentally, $\eta_{\rm int}$ reaches a maximum of 93\% at a pump power of 52 mW (Fig.~\ref{Fig:group_figure2}(b)). Numerical simulations, incorporating measured propagation losses ($\alpha_{1}=0.22$, $\alpha_{2}=0.20$, $\alpha_{3}=0.12~\mathrm{cm}^{-1}$), predict a peak of 111\%---a value exceeding unity because the signal experiences higher loss than the DFG photons. The discrepancy between the measured and simulated maxima is caused by higher-order 393-nm modes in the waveguide that are not phase-matched, leading to an underestimation of the fundamental guided mode efficiency. In the low-conversion regime, $\eta_{\rm int}$ is related to $\eta_{\rm nor}$ via~\cite{bortz2002IEEE-propagationloss,langrock2007classical}:
\begin{equation}\eta_{\rm int} = \frac{\eta_{\rm nor} P_2^{\rm out}(e^{\Delta\alpha L}-1)^2}{(\Delta\alpha)^2}, \quad\Delta\alpha = \frac{\alpha_1+\alpha_2-\alpha_3}{2}.
\label{Eq:efficiency with loss}
\end{equation}
Fitting the data to Eq.~(\ref{Eq:efficiency with loss}) yields $\eta_{\rm nor} = 703 \pm 17\ \%\rm/(W\cdot cm^2)(95\% \rm ~confidence~interval)$. Correcting for the higher-mode contribution revises the normalized efficiency of the fundamental guided mode to $839 \pm 20\ \%\rm/(W\cdot cm^2)$, which is in excellent agreement with the ideal prediction of 831\%/(W·c\(\mathrm{m^{2}}\)) (see Supplemental Material for details).


\begin{figure}[htbp]
    \centering  \includegraphics[width=1\linewidth]{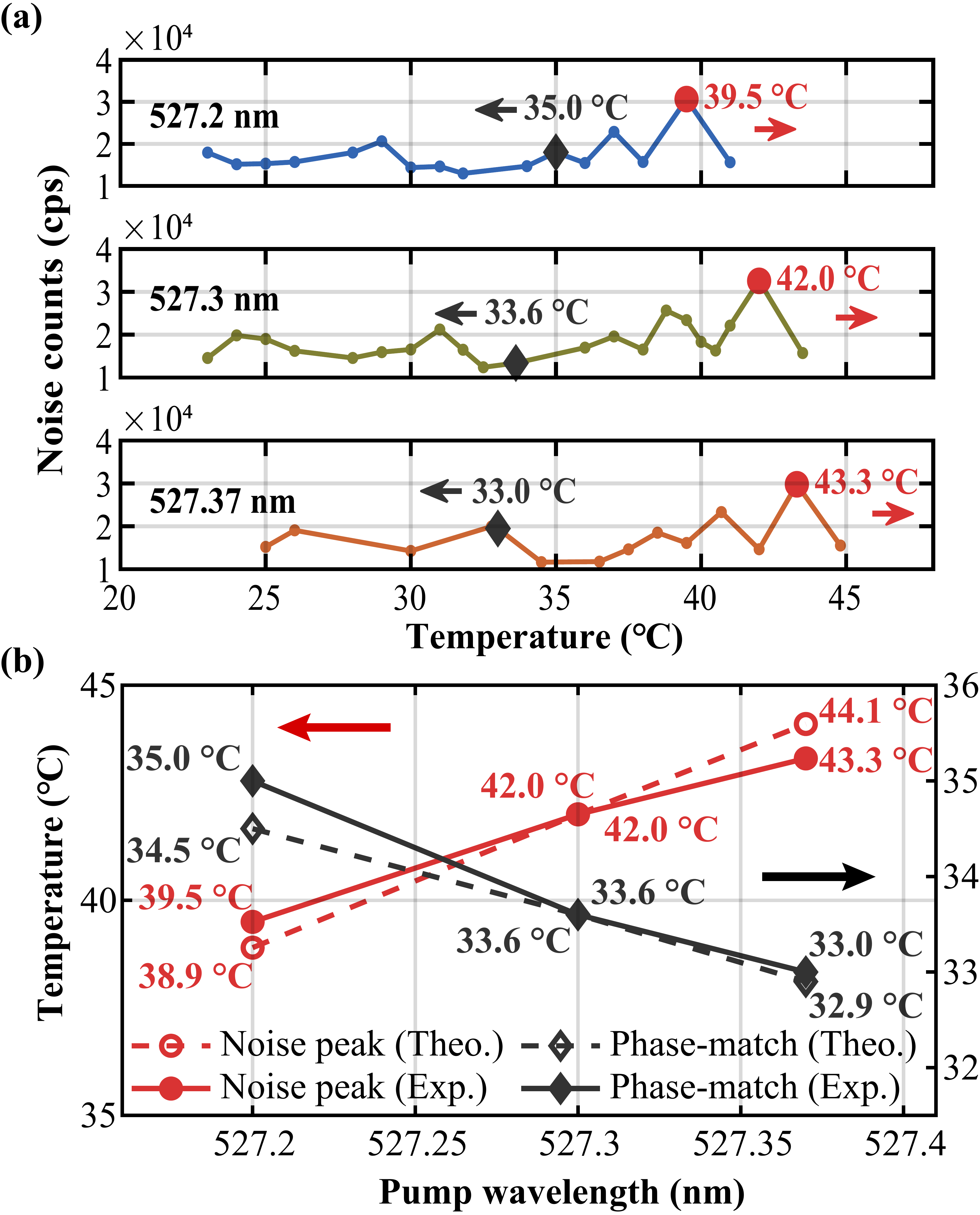}
    \caption{(a) Noise counts versus waveguide temperature at a fixed pump power for pump wavelengths of 527.2 nm (blue), 527.3 nm (olive), and 527.37 nm (orange-red). Filled red circles and black diamonds mark the noise counts and corresponding waveguide temperatures at the noise peaks and DFG phase-matching points, respectively. Arrows indicate the opposing temperature shifts versus pump wavelength. (b) Temperatures of the noise peaks and DFG phase-matching points labeled in (a) versus pump wavelength. Filled and open red circles denote theoretical and experimental noise peaks, respectively; filled and open black diamonds denote theoretical and experimental DFG phase-matching points, respectively.}
    \label{Fig.Noise_vs_temperature}
\end{figure}

\textbf{\textit{Noise Behavior and Suppression.}}---Prior studies have attributed the noise in short-wavelength pumping to spontaneous Raman scattering and spontaneous parametric down-conversion (SPDC)~\cite{mann2024-oe-short-wavelength-pump}. However, effective SPDC noise suppression methods are lacking, leaving spectral filtering as the predominant mitigation strategy. Although achieving theoretical $\eta_{\rm nor}$ establishes a low-noise baseline by minimizing the required pump power, the residual noise remains non-negligible for practical quantum networking. To address this, we investigate the noise behavior to develop more robust suppression strategies.


Theoretically, the DFG and SPDC processes---driven by the 527-nm pump---are predicted to exhibit distinct phase-matching responses. By modeling the SPDC noise as a first-order QPM process as shown in Fig.~\ref{Fig.Noise_vs_temperature}(b), the QPM-temperatures versus pump wavelengths for DFG and SPDC are found to shift at rates of approximately $-0.01$ $^\circ\mathrm{C}/\mathrm{pm}$ and $+0.03$ $^\circ\mathrm{C}/\mathrm{pm}$, respectively. 
This divergence reveals that the signal and noise processes shift in opposite directions and at significantly different rates with respect to the pump wavelength.


\begin{figure*}[htbp]
    \centering
    \includegraphics[width=1\linewidth]{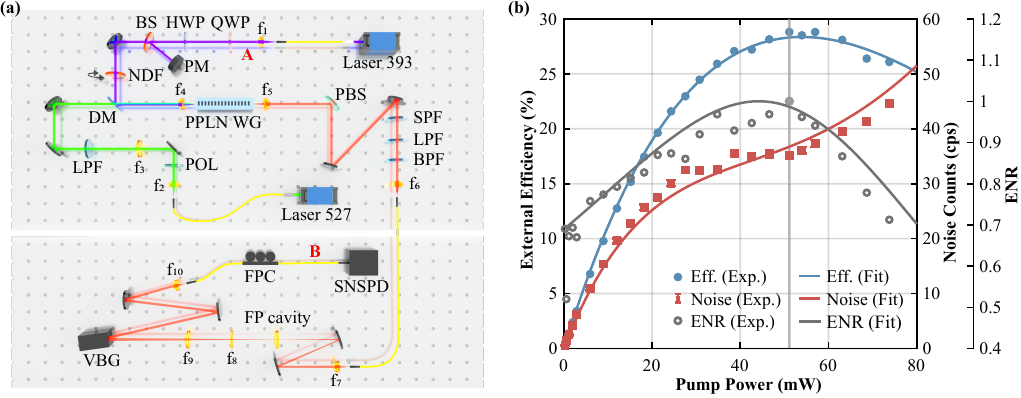}
    \caption{(a) Experimental setup. The upper section illustrates the DFG module. QWP: quarter-wave plate, HWP: half-wave plate, BS: beam splitter, PM: power meter, NDF: neutral density filter, DM: dichroic mirror, POL: polarizer, LPF: long-pass filter, SPF: short-pass filter, BPF: band-pass filter, PBS: polarizing beam splitter, PPLN WG: periodically poled lithium niobate waveguide. The lower section shows the ultra-narrowband filtering stage. FP: Fabry--P\'erot, VBG: volume Bragg grating, FPC: fiber polarization controller, SNSPD: superconducting nanowire single-photon detector. (b) External efficiency and noise counts of the DFG process versus pump power. Filled blue circles show the measured external efficiency, with the solid blue line representing the fitted $\rm sin^{2}$-type model. Filled orange-red squares denote the measured noise counts, and the dashed red lines show the fits based on the noise model [Eq.~(\ref{Eq:modified-noise})] including propagation losses. Open gray circles indicate the measured external efficiency-to-noise ratio (ENR), with the solid green line showing the fitted trend. The gray vertical line marks the operating point during the experiment.}
    \label{Fig.External Efficiency DC}
\end{figure*}

We experimentally characterized the noise using an InGaAs detector with a 0.8 nm spectral bandwidth to avoid statistical fluctuations from low counts after subsequent ultra-narrowband filtering. Figure~\ref{Fig.Noise_vs_temperature}(a) illustrates the noise dependence on waveguide temperature and pump wavelength. The measured noise exhibits pronounced temperature-dependent fluctuations and an overall shift as the pump wavelength is tuned from 527.2 nm to 527.3 nm and 527.37 nm. Crucially, the evolutions of the DFG phase-matching temperature and the noise peak locations (marked in Fig.~\ref{Fig.Noise_vs_temperature}(a)) diverge as expected. As detailed in Fig.~\ref{Fig.Noise_vs_temperature}(b), these experimental shifts (DFG: $-0.01$ $^\circ\mathrm{C}/\mathrm{pm}$, SPDC: $+0.02$ $^\circ\mathrm{C}/\mathrm{pm}$) closely follow the theoretical trend, confirming the presence of pronounced SPDC-induced noise driven by the 527-nm pump.


Therefore, by exploiting the opposite tuning characteristics of DFG and SPDC, we propose a noise suppression method: fine-tuning of the pump wavelength can separate the phase-matching point from the noise peak, thereby reducing the nonlinear noise. As illustrated in Fig.~\ref{Fig.Noise_vs_temperature}(a), for a 527.37-nm pump, the phase-matching temperature of $33.0^{\circ}\mathrm{C}$ coincides with a local noise peak, while tuning the pump to 527.30 nm shifts this temperature to $33.6^{\circ}\mathrm{C}$, near a local noise minimum. Moreover, a broader pump wavelength tuning range enables shifting the DFG operating point from a noise maximum to a minimum, yielding a threefold reduction in noise. 


\textbf{\textit{QFC Module Performance.}}---Enabled by the high-performance PPLN waveguide and the effective noise suppression mechanism, we constructed a QFC module to characterize the system performance(Fig.~\ref{Fig:group_figure2}(a)). Continuous-wave lasers at 393 nm (signal) and 527 nm (pump) are polarization-adjusted for type-0 phase matching before being coupled into the ridge waveguide. The Gaussian beam parameters of the signal and pump are individually optimized to compensate for chromatic dispersion in the coupling lens, ensuring optimal coupling for both beams simultaneously. Aided by the anti-reflection coatings on the waveguide facets, the transmission reaches 49\% ($T_{\rm WG}$) and 60\% for the signal and pump, respectively. 


At the waveguide output, a primary noise suppression stage is implemented in free space prior to coupling the photons into a single-mode fiber (SMF-28e). The first filtering stage comprises two long-pass filters (cutoff at 1500 nm), one short-pass filter (cutoff at 1600 nm), a 1550 nm band-pass filter with 0.8-nm full width at half maximum (FWHM), and a reflective polarizing plate beam splitter to reject orthogonally polarized noise, which constitutes $\sim$20--30\% of the total noise~\cite{zhang2020two-noise-pol}. Following this pre-filtering, the generated 1550 nm photons are collected into the SMF-28e fiber with a combined efficiency of $80\%$ ($T_{\rm collect}$). Subsequently, an ultra-narrowband filtering stage is employed to further reduce residual noise. It consists of a Fabry--P\'erot cavity with a 40-MHz FWHM and 45-GHz free spectral range, combined with a 10-GHz FWHM volume Bragg grating to ensure single-peak transmission. This stage provides an overall transmission of 79\% ($T_{\rm filter}$). After filtering, the photons are detected by a superconducting nanowire single-photon detector with a detection efficiency of 85\%.

Based on this configuration, we characterized the overall performance of this module (from point A to B in Fig.~\ref{Fig.External Efficiency DC}(a)). For accurate efficiency characterization, a flippable beam splitter is used to simultaneously monitor the input signal and generated DFG photons. The external efficiency $\eta_{\rm ext}$ is expressed as
\begin{equation}
    \eta_{\rm ext} = T_{\rm WG} \times \eta_{\rm int} \times T_{\rm collect} \times T_{\rm filter}.
    \label{Eq.eta_external}
\end{equation}
Evaluated via Eq.~\ref{Eq.eta_external}, the module achieves a maximum $\eta_{\rm ext}$ of 28.8\% at a pump power of 52 mW, while maintaining a low noise count of 35 cps (Fig.~\ref{Fig.External Efficiency DC}(b)). Compared to previous demonstrations listed in Table~\ref{Tab:comparison}, this performance represents a major breakthrough for UV QFC, enabling practical, on-site configurable deployment.

\begin{table}[t!]
\centering
\caption{Performance Comparison of UV and Blue Region to Telecom Band Conversion.}
\setlength{\tabcolsep}{2mm}
\begin{tabular}{lccc}
\hline\hline
Reference & Signal & Efficiency & Noise \\
\midrule
Kasture et al.~\cite{kasture2016JOP-UVQFC-Yb369->1580} & 369 nm & 0.02\% & $\sim10^{6}$ cps \\
Yu et al.~\cite{yu2025APL-UVQFC-Yb369->456->864->1552} & 369 nm & ... & ... \\
Wright et al.~\cite{wright2018PRA-UVQFC-Sr425->1560-1548->422} & 426 nm & 1.1\% & ... \\
This work & 393 nm & 28.8\% & 35 cps \\
\hline\hline
\end{tabular}
\label{Tab:comparison}
\end{table}
Notably, the external efficiency-to-noise ratio (ENR) coincides with maximum conversion efficiency (Fig.~\ref{Fig.External Efficiency DC}(b)), in contrast to long-wavelength schemes where the ENR maximum typically precedes the efficiency peak~\cite{weinfurter2020prl-atom-780-20km}. This feature arises from the noise saturation observed at the maximum conversion efficiency. While a model~\cite{maring2018Optica-short-wavelength-pump-noise} based on the back-conversion of pump-induced noise explains the saturation mechanism, the predicted plateau significantly lags behind the experimental values (see Supplemental Material). We attribute this deviation to propagation loss, and propose a refined model:
\begin{equation}
    \begin{aligned}
        N(P) &=  a\cdot  P \int_0^L e^{\alpha_{\rm pump}(L- x)}  
        \Big[ 1 - \Big(1-e^{-\alpha_{\rm DFG}(L- x)}\Big) \\&-\eta_{\rm int}^{\rm max}
        \sin^2 \Big( (L - x) \sqrt{\eta_{\rm nor} P e^{\alpha_{\rm pump}(L- x)}} \Big) \Big] dx,
    \end{aligned}
    \label{Eq:modified-noise}
\end{equation}
where $a$ (Hz/W/cm) is the linear noise generation coefficient. Fitted with measured losses $\alpha_{i}$, Eq.~(\ref{Eq:modified-noise}) accurately reproduces the experimental noise plateau, confirming the critical role of propagation loss in shaping the distinct noise-efficiency characteristics of short-wavelength QFC (see Supplemental Material for more details).

\textbf{\textit{Conclusion.}}---In this letter, we address the critical challenges in high-performance UV-to-telecom QFC via pushing efficiency and noise performance to their theoretical limit. Through modeling and experimental verification of the impact of domain defects on normalized conversion efficiency, we establish a quantitative threshold for defect counts and realize the maximum achievable $\eta_{\rm nor}$. Moreover, we propose a noise suppression mechanism exploiting the counter-tuning behaviors of the DFG and SPDC processes, enabling a 3-fold noise reduction. We further refined the noise-pump model to incorporate propagation losses, enabling accurate modeling of the noise dynamics in the QFC process. Combining these advances with optimized coupling and ultra-narrowband filtering, we demonstrate a 393 nm-to-1550 nm converter achieving a record-high system efficiency of 28.8\% and an ultra-low noise of 35 counts per second. This performance sets a new benchmark for UV QFC and provides a critical interface for short-wavelength quantum memories, paving the way for advancements in long-distance quantum communication, distributed quantum computing, and quantum sensing.



Further improvement of fabrication process to reduce the propagation loss of PPLN waveguide promises further efficiency gains. The methodology presented here is readily extendable to other critical ion-trap transitions, including 369 nm ($^{171}$Yb$^+$), 422 nm ($^{88}$Sr$^+$), and 493 nm ($^{138}$Ba$^+$). For these systems, we anticipate further enhanced overall efficiencies at longer wavelengths, benefiting from reduced material absorption. Importantly, the strategy for reaching the theoretical efficiency limit is universally applicable to generic QFC process, unlocking the potential of hybrid quantum networks. 

\begin{acknowledgments}
\textbf{\textit{Acknowledgments.}}---
This work has been supported by the National Natural Science Foundation of China (Grant No. T2125010), the Quantum Science and Technology-National Science and Technology Major Project (Grant No.2021ZD0300800, No.2021ZD0300802), the Key R$\&$D Program of Shandong Province, China (Grant No. 2024CXPT083),  Natural Science Foundation of Shandong Province (Grant No. ZR2021LLZ013, No. ZR2022LLZ009, No. ZR2022LLZ010, No. ZR2023LLZ006), Shandong Postdoctoral Science Foundation (Grant No. SDCX-ZG-202400330), Q. Z. were supported by the New Cornerstone Science Foundation through the Xplorer Prize, the SAICT Experts Program, the Taishan Scholar Program of Shandong Province, and Quancheng Industrial Experts Program. M.-Y. Z. were supported by the Taishan Scholar Program of Shandong Province and Haiyou Plan Project of Jinan.
\end{acknowledgments}

\bibliography{DFG393to1550}
\end{document}


\title{High-Performance Quantum Frequency Conversion Connecting Ultraviolet and Telcom Bands: Supplemental Material}

\author{Yi Yang}
\affiliation{Hefei National Research Center for Physical Sciences at the Microscale and School of Physical Sciences, University of Science and Technology of China, Hefei 230026, China.}
\affiliation{Jinan Institute of Quantum Technology and CAS Center for Excellence in Quantum Information and Quantum Physics, University of Science and Technology of China, Jinan 250101, China.}

\author{Bin Wang}
\affiliation{Jinan Institute of Quantum Technology and CAS Center for Excellence in Quantum Information and Quantum Physics, University of Science and Technology of China, Jinan 250101, China.}

\author{Ji-Chao Lin}
\author{Yang Gao}
\author{Xin Li}
\affiliation{Jinan Institute of Quantum Technology and CAS Center for Excellence in Quantum Information and Quantum Physics, University of Science and Technology of China, Jinan 250101, China.}

\author{Jiu-Peng Chen}
\affiliation{Hefei National Laboratory, University of Science and Technology of China, Hefei 230088, China.}
\affiliation{Jinan Institute of Quantum Technology and CAS Center for Excellence in Quantum Information and Quantum Physics, University of Science and Technology of China, Jinan 250101, China.}

\author{Lei Hou}
\affiliation{Hefei National Laboratory, University of Science and Technology of China, Hefei 230088, China.}
\affiliation{Shanghai Research Center for Quantum Science and CAS Center for Excellence in Quantum Information and Quantum Physics, University of Science and Technology of China, Shanghai 201315, China.}

\author{Ye Wang}
\author{Yong Wan}
\affiliation{Hefei National Research Center for Physical Sciences at the Microscale and School of Physical Sciences, University of Science and Technology of China, Hefei 230026, China.}
\affiliation{Hefei National Laboratory, University of Science and Technology of China, Hefei 230088, China.}
\affiliation{Shanghai Research Center for Quantum Science and CAS Center for Excellence in Quantum Information and Quantum Physics, University of Science and Technology of China, Shanghai 201315, China.}

\author{Xiu-Ping Xie}
\author{Ming-Yang Zheng}
\affiliation{Jinan Institute of Quantum Technology and CAS Center for Excellence in Quantum Information and Quantum Physics, University of Science and Technology of China, Jinan 250101, China.}
\affiliation{Hefei National Laboratory, University of Science and Technology of China, Hefei 230088, China.}

\author{Qiang Zhang}
\affiliation{Hefei National Research Center for Physical Sciences at the Microscale and School of Physical Sciences, University of Science and Technology of China, Hefei 230026, China.}
\affiliation{Jinan Institute of Quantum Technology and CAS Center for Excellence in Quantum Information and Quantum Physics, University of Science and Technology of China, Jinan 250101, China.}
\affiliation{Hefei National Laboratory, University of Science and Technology of China, Hefei 230088, China.}
\affiliation{Shanghai Research Center for Quantum Science and CAS Center for Excellence in Quantum Information and Quantum Physics, University of Science and Technology of China, Shanghai 201315, China.}

\author{Jian-Wei Pan}
\affiliation{Hefei National Research Center for Physical Sciences at the Microscale and School of Physical Sciences, University of Science and Technology of China, Hefei 230026, China.}
\affiliation{Hefei National Laboratory, University of Science and Technology of China, Hefei 230088, China.}
\affiliation{Shanghai Research Center for Quantum Science and CAS Center for Excellence in Quantum Information and Quantum Physics, University of Science and Technology of China, Shanghai 201315, China.}

\date{\today}

\begin{abstract}
\end{abstract}

\maketitle

\section*{Impact of Domain Defects on Conversion Efficiency}
\begin{figure}[htbp]
    \centering
    \includegraphics[width=0.8\linewidth]{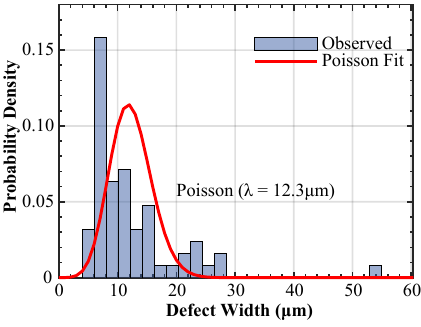}
    \caption{Defect width distribution. The histogram displays the experimental data, while the solid red curve represents the Poisson distribution defined by the mean width.}
    \label{Fig.defect_width_poisson}
\end{figure} 
\begin{figure*}[htbp]
    \centering
    \includegraphics[width=1\linewidth]{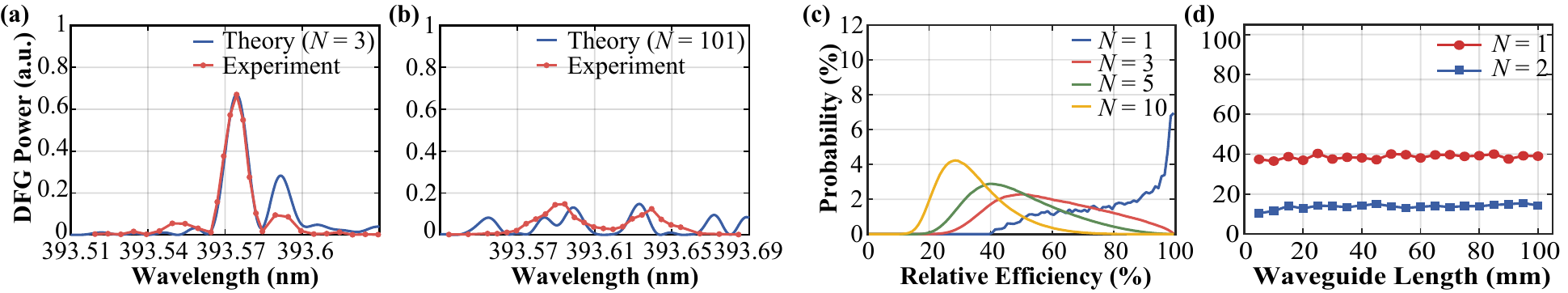}
    \caption{Analysis of domain defect impacts on spectral response and relative conversion efficiency. (a) and (b) Comparison between experimental (orange solid line, filled circles) and theoretical (blue solid lines) spectral tuning curves for waveguides containing (a) $N=3$ and (b) $N=101$ domain defects. (c) Simulated probability distribution of the relative normalized efficiency for varying defect counts. (d) Probability of achieving $>90\%$ relative efficiency as a function of waveguide length for $N=1$ (red solid line, filled circles) and $N=2$ (blue solid line, filled squares).}
    \label{Fig.Analysis of domain defect impacts}
\end{figure*} 
Beyond the efficiency degradation discussed in the main text, random phase shifts induced by domain inversion defects cause significant distortions in the spectral tuning curve. Theoretically, this response is determined by the Fourier transform of the nonlinear coefficient distribution $d(z)$: $\eta \propto \left| \int_{0}^{L} d(z) e^{-i 2\pi q z} dz \right|^2$, where $d(z)$ is defined in Equations (1) and (2) of the main text.

To validate the defect model, we characterize two 20-mm-long waveguides containing 3 and 101 domain defects, respectively. Figure~\ref{Fig.Analysis of domain defect impacts}(a-b) presents the theoretical tuning curves calculated using the measured defect widths and distributions. The theoretical predictions capture the overall spectral trend of the experimental data. However, discrepancies are observed, particularly in the waveguide with a higher defect count (101 defects). We attribute these deviations to simplifications in the model, which neglects structural non-uniformities such as variations in film thickness and waveguide geometry, combined with measurement errors inherent to the finite resolution of microscopic inspection. Despite these deviations, the substantial agreement between theory and experiment confirms that domain defects act as the dominant mechanism limiting the normalized conversion efficiency. 

Building upon the validated model, we further investigate the impact of domain defects via simulations, assuming randomly distributed defect positions and experimentally derived Poisson-distributed widths (Fig.~\ref{Fig.defect_width_poisson}) ($\lambda \sim 12.3\,\mu\text{m}$). Figure~\ref{Fig.Analysis of domain defect impacts}(c) illustrates the probability distribution of the relative conversion efficiency for varying defect counts. As the defect count increases, the distribution exhibits a pronounced peak shift toward the lower efficiency regime, indicating a rapidly declining probability of achieving the theoretical normalized efficiency. Quantitatively, the probability of maintaining a relative efficiency exceeding 90\% drops to approximately 40\% for a single defect ($N=1$) and only 15\% for two defects ($N=2$), as shown in Fig~\ref{Fig.Analysis of domain defect impacts}(d). Furthermore, the simulations in figure~\ref{Fig.Analysis of domain defect impacts}(d) suggest that for centimeter-scale waveguides, the impact of defect quantity on conversion efficiency remains consistent regardless of the specific waveguide length. Consequently, realizing near-theoretical normalized efficiency necessitates strictly limiting the number of domain defects to $N \le 2$.

\section*{Experimental Characterization of Normalized Efficiency}
\subsection{Propagation Loss Characterization}
\begin{figure}[htbp]
    \centering
    \includegraphics[width=1\linewidth]{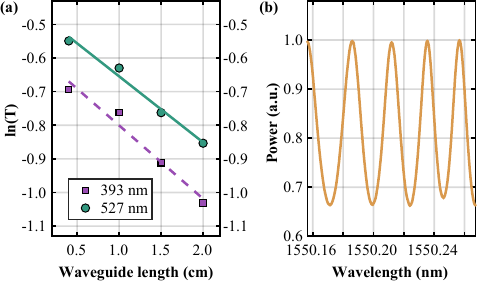}
    \caption{Transmission loss of the periodically poled lithium niobate waveguide. (a) Cut-back measurements at 393 and 527 nm, where $T$ denotes the waveguide transmission. Linear fits of $\ln(T)$ versus length yield the propagation. (b) Fabry--Pérot measurements at 1550 nm, showing normalized transmission spectrum; propagation loss is extracted from the fringe peak-to-valley contrast.}
    \label{Fig.propagation loss}
\end{figure}
Propagation losses are measured on periodically poled lithium niobate waveguides at 393 nm, 527 nm, and 1550 nm (Fig.~\ref{Fig.propagation loss}). At 393 nm and 527 nm, these losses are determined using the cut-back method~\cite{kappeler2011-cutback}. We record the transmission $T$ of waveguides with varying lengths and extract the propagation loss from the slope of linear fits of $\ln(T)$ versus length. This analysis yields attenuation coefficients of $\alpha_1 = 0.22~\rm cm^{-1}$ and $\alpha_2 = 0.20~\rm cm^{-1}$ at 393 nm and 527 nm, respectively. 

At 1550 nm, we determine the propagation loss via a Fabry--P\'erot interferometric method~\cite{taebi2008-fp-interferometric}. The normalized transmission spectrum exhibits interference fringes. The propagation loss is then calculated from the fringe peak-to-valley contrast $b$ according to:
\begin{equation}
\alpha = \frac{1}{L} \ln \frac{R}{\bar{R}}, \quad
R = \left(\frac{n-1}{n+1}\right)^2, \quad
\bar{R} = \frac{1-\sqrt{b}}{1+\sqrt{b}},
\end{equation}
where $R$ is the Fresnel reflection at the waveguide facets, $\bar{R}$ is derived from the measured peak-to-valley contrast, $n$ is the waveguide refractive index at 1550 nm, and $L$ is the waveguide length. Using this formula, the propagation loss at 1550 nm is calculated to be $\alpha_3 = 0.12~\rm cm^{-1}$.
\subsection{Propagation-loss-corrected Normalized Efficiency}
\begin{figure}[htbp]
    \centering
    \includegraphics[width=0.9\linewidth]{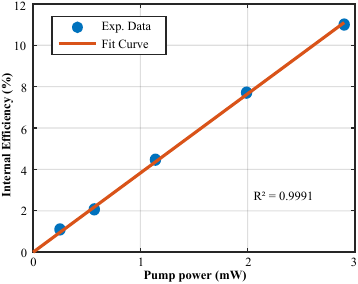}
    \caption{Fit of the normalized conversion efficiency using the first five low-conversion points from Fig.~2(b). Filled circles represent the measured internal efficiency, and the solid line shows the linear fit.}
    \label{Fig.Normalized_Efficiency}
\end{figure}
Under the undepleted pump approximation and negligible propagation loss, the relationship between $\eta_{\rm int}$ and $\eta_{\rm nor}$ is commonly expressed as
\begin{equation}
    \label{Eq.eta_relation_sin2}
    \eta_{\rm int} = \sin^2\bigl(\sqrt{\eta_{\rm nor} P_2} L\bigr),
\end{equation}
with $L$ denoting the waveguide length. However, significant propagation losses, particularly in the ultraviolet wavelength range, lead to deviations from ideal behavior. To account for this, we adopt a more realistic efficiency model that includes propagation losses, in which the coupled-mode equations under phase-matching conditions take the form~\cite{bortz2002IEEE-propagationloss,roussev2007PhD-propagationloss}:
\begin{equation}
    \begin{cases}
        \displaystyle
        \frac{d A_{1}}{dz} = -i \kappa_{1} \theta^{*} A_{3} A_{2}^{*} - \frac{\alpha_{1}}{2} A_{1}, \\[12pt]
        \displaystyle
        \frac{d A_{2}}{dz} = -i \kappa_{2} \theta^{*} A_{3} A_{1}^{*} - \frac{\alpha_{2}}{2} A_{2}, \\[12pt]
        \displaystyle
        \frac{d A_{3}}{dz} = -i \kappa_{3} \theta A_{1} A_{2} - \frac{\alpha_{3}}{2} A_{3},
    \label{Eq.coupled-mode equations with loss}
    \end{cases}
\end{equation}
where $A_i$, $\alpha_i$, $\theta$, and $\kappa_i$ denote the field amplitude, power attenuation coefficient, mode overlap integral, and normalization constant of wave $i$, respectively. Under the low-conversion-efficiency approximation ($-i \kappa_{1} \theta^{*} A_{3} A_{2}^{*}=0$, $-i \kappa_{2} \theta^{*} A_{3} A_{1}^{*}=0$), the corrected internal conversion efficiency can be expressed as~\cite{bortz2002IEEE-propagationloss,roussev2007PhD-propagationloss}
\begin{equation}
    \eta_{\rm int}=\frac{\eta_{\text{nor}}P^{\text{out}}_{2}(e^{\Delta\alpha L}-1)^2}{{(\Delta\alpha)}^2}
    \label{Eq:efficiency with loss}
\end{equation}
with $\Delta\alpha=(\alpha_1+\alpha_2-\alpha_3)/2$. 

Fitting the low-conversion data from Fig.~2(b) of the main text to Eq.~(\ref{Eq:efficiency with loss}) yields a normalized conversion efficiency of $703 \pm 17\ \%\rm/(W\cdot cm^2)$ ($95\% \rm ~confidence~interval,~R^2 = 0.9991$), as shown in Fig.~\ref{Fig.Normalized_Efficiency}. By contrast, the standard lossless $\sin^2$-type expression (Eq.~(\ref{Eq.eta_relation_sin2})) yields 1141\%/(W·cm$^2$), significantly overestimating the theoretical value.

\section*{Refinement of the Noise-Power Model}
\begin{figure}[htbp]
    \centering
    \includegraphics[width=1\linewidth]{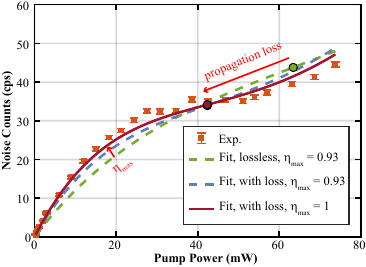}
    \caption{Noise counts as a function of pump power. Orange-red squares: measured data. Green and blue dashed lines: noise fits without and with propagation losses, respectively, both assuming a maximum conversion efficiency of 93\%. Orange-red solid line: noise fit including propagation losses and assuming a 100\% maximum efficiency.}
    \label{Fig.Noise DC loss no loss}
\end{figure}
According to the noise mechanism described in Ref.~\cite{maring2018Optica-short-wavelength-pump-noise}, the pump-power dependence of the noise induced by the short-wavelength pump is given by:
\begin{equation}
    \begin{aligned}
        N(P) =  a\cdot P \int_0^L 
        \big( 1 - \eta_{\rm int}^{\rm max}
        \sin^2\!\big((L-x)\sqrt{\eta_{\rm nor}P}\big) dx,
    \end{aligned}
    \label{Eq:noloss-noise}
\end{equation}
where $a$ (Hz/W/cm) is the linear noise generation coefficient and $L$ is the waveguide length. Fitting our experimental data with this lossless model reveals a clear deviation: the predicted flattest point of noise growth occurs at a higher pump power than observed (Fig.~\ref{Fig.Noise DC loss no loss}). To accurately reproduce the experimental behavior, we identify two critical corrections: propagation loss and the maximum internal conversion efficiency. Incorporating these factors yields the modified expression:
\begin{equation}
    \begin{aligned}
        N(P) &=  a\cdot  P \int_0^L e^{\alpha_{\rm pump}(L- x)}  
        \Big[ 1 - \Big(1-e^{-\alpha_{\rm DFG}(L- x)}\Big) \\&-\eta_{\rm int}^{\rm max}
        \sin^2 \Big( (L - x) \sqrt{\eta_{\rm nor} P e^{\alpha_{\rm pump}(L- x)}} \Big) \Big] dx.
    \end{aligned}
    \label{Eq:modified-noise}
\end{equation}
The comparison of the experimental data fitted according to models with and without correction is shown in Figure~\ref{Fig.Noise DC loss no loss}. Two key improvements are evident. First, propagation loss accounts for the shift in saturation power, reproducing the observed localized plateau. Second, adjusting $\eta_{\rm int}^{\rm max}$ from the measured 93\% to the theoretical 100\% yields a fit closer to the experimental curvature, as the back-conversion of noise in the telecom band is not constrained by higher-order modes. Collectively, accounting for both propagation loss and efficiency correction enables precise theoretical prediction of noise behavior in QFC processes.

\section*{Long-term System Stability}
\begin{figure}[htbp]
    \centering
    \includegraphics[width=1\linewidth]{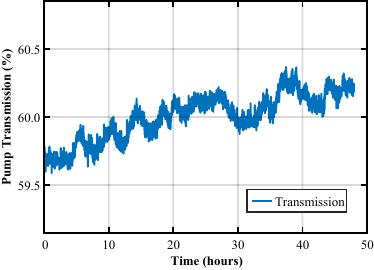}
    \caption{Long-term monitoring of pump transmission into the waveguide.}
    \label{Fig.setup_stability}
\end{figure}
The difference-frequency generation setup exhibits excellent long-term stability, maintained through precise waveguide temperature control (peak-to-peak fluctuation $<5$ mK) and an enclosed optical path that minimizes air currents and ambient temperature drifts. Under these conditions, pump coupling into the waveguide is continuously monitored for 48 h, showing a peak-to-peak relative fluctuation of 1.4$\%$ and an RMS fluctuation of 0.28$\%$(Fig.~\ref{Fig.setup_stability}). The Fabry--P\'erot cavity is actively stabilized to 1 mK peak-to-peak, corresponding to $\sim$1.2 MHz frequency drift. These measures ensure system stability suitable for long-duration operation in quantum network applications.
\begin{acknowledgments}
\end{acknowledgments}

\bibliography{DFG393to1550}